\documentclass[12pt,twoside, a4paper]{article}
\def\pd{\partial}
\def\mc{\mathcal}

\usepackage[dvips]{graphicx}
\usepackage{amssymb}
\usepackage{amssymb,amsmath}
\usepackage{graphicx}
\input{epsf.sty}  
\begin{document}
\begin{center}
\LARGE{\textbf{Supersymmetric Janus solutions in four-dimensional
$N=3$ gauged supergravity}}
\end{center}
\vspace{1 cm}
\begin{center}
\large{\textbf{Parinya Karndumri}}
\end{center}
\begin{center}
String Theory and Supergravity Group, Department
of Physics, Faculty of Science, Chulalongkorn University, 254 Phayathai Road, Pathumwan, Bangkok 10330, Thailand
\end{center}
E-mail: parinya.ka@hotmail.com \vspace{1 cm}\\
\begin{abstract}
We construct supersymmetric Janus solutions using four-dimensional
$N=3$ gauged supergravity with $SO(3)\times SU(3)$ gauge group. The
$N=3$ supersymmetric $AdS_4$ vacuum with unbroken $SO(3)\times
SU(3)$, identified with the compactification of eleven-dimensional
supergravity on $AdS_4\times N^{010}$, provides a gravity dual of
supersymmetric $N=3$ Chern-Simons-Matter theory in three dimensions
with $SU(3)$ flavor symmetry. The Janus solutions accordingly
describe supersymmetric conformal interfaces within this
Chern-Simons-Matter theory via the AdS/CFT holography. We find two
classes of Janus solutions preserving respectively $(2,1)$ and
$(0,1)$ supersymmetry on the $(1+1)$-dimensional conformal defects.
The solution with $(2,1)$ supersymmetry preserves $SO(2)\times
SO(2)\times SO(2)\subset SO(3)\times SU(3)$ symmetry while the
$(0,1)$ supersymmetric solution is invariant under a larger
$SO(2)\times SU(2)\times SO(2)$ symmetry.
\end{abstract}
\newpage
\section{Introduction}
Conformal field theories (CFTs) with a conformal defect are
important in describing some properties of condensed matter and
statistical physics systems \cite{Cardy_defect,dCFT}. The AdS/CFT
correspondence \cite{maldacena} offers holographic duals to these
theories via Janus solutions \cite{Bak_Janus}. According to the
usual relations in the AdS/CFT correspondence, these solutions would
be useful in studying strongly coupled CFT with conformal
interfaces. Along this line, holographic duals of conformal defects
within $N=4$ supersymmetric Yang-Mills (SYM) theory, one of the
primary examples of the AdS/CFT duality, have been extensively
studied in a number of previous works, see for example
\cite{Freedman_Janus,DHoker_Janus,Witten_Janus,Freedman_Holographic_dCFT}.
\\
\indent In general, Janus solutions can be obtained from AdS-sliced
domain walls. A special case of flat domain walls with the
$AdS$-slice replaced by the flat Minkowski space describes the usual
holographic RG flows. Supersymmetric Janus solutions in five
dimensions, dual to $N=4$ SYM with conformal interfaces, have been
obtained both from five-dimensional gauged supergravity and directly
from ten-dimensional type IIB theory
\cite{5D_Janus_CK,5D_Janus_DHoker1,5D_Janus_DHoker2,5D_Janus_Suh}.
Janus solutions dual to defect conformal field theories (dCFTs) or
interface conformal field theories (ICFTs) in two dimensions have
also been studied in
\cite{3D_Janus_de_Boer,3D_Janus_Bachas,3D_Janus_Bak,half_BPS_AdS3_S3_ICFT,exact_half_BPS_string}
including the multi-face Janus solution recently constructed in
\cite{multi_face_Janus}.
\\
\indent In four dimensions, a class of supersymmetric Janus
solutions from eleven-dimensional M-theory have been classified in
\cite{4D_Janus_from_11D}. A number of supersymmetric Janus solutions
in the maximal $N=8$ gauged supergravity with various symmetries
have been studied in \cite{warner_Janus}. The solution with
$SO(4)\times SO(4)$ symmetry can be uplifted to eleven dimensions
and has been shown to be different from the solutions classified in
\cite{4D_Janus_from_11D}. All solutions considered in
\cite{warner_Janus} have been obtained by truncating the
$E_{7(7)}/SU(8)$ scalar manifold of the $N=8$ supergravity to a
single complex scalar living in $SL(2,\mathbb{R})/SO(2)$ coset.
\\
\indent In this paper, we will give another example of
supersymmetric Janus solutions within $N=3$ gauged supergravity
coupled to eight vector multiplets. This results in $N=3$ gauged
supergravity with $SO(3)\times SU(3)$ gauge group which is expected
to arise from a dimensional reduction of eleven-dimensional
supergravity on a tri-sasakian manifold $N^{010}$
\cite{Castellani_Romans,N3_spectrum1,N3_spectrum2}. Possible
three-dimensional $N=3$ SCFT dual to the supersymmetric $AdS_4$
critical point with $SO(3)\times SU(3)$ symmetry has been given in
\cite{Ring_N3_superfield,Shadow_N3_multiplet}. Other $AdS_4$
critical points and holographic RG flows between $AdS_4$ critical
points including flows to $AdS_2$ geometries have been extensively
studied in \cite{N3_SU2_SU3}.
\\
\indent We will consider supersymmetric Janus solutions in this
$N=3$ gauged supergravity with $SO(2)\times SU(2)\times SO(2)$ and
$SO(2)\times SO(2)\times SO(2)$ symmetries. As we will see, the
solutions preserve respectively $N=(1,0)$ and $N=(2,1)$
supersymmetries on the $(1+1)$-dimensional interfaces. According to
the AdS/CFT correspondence, these solutions should be dual to some
conformal defects, breaking $N=3$ supersymmetry and $SO(3)\times
SU(3)$ global symmetry, in the Chern-Simons-Matter theory dual to
the $AdS_4\times N^{010}$ background.
\\
\indent The paper is organized as follow. In section \ref{N3theory},
we review the matter-coupled $N=3$ gauged supergravity in order to
set up the notations and collect all the needed formulae. The
analysis of BPS equations relevant to finding supersymmetric Janus
solutions will also be given. Supersymmetric Janus solutions with
$N=(1,0)$ supersymmetry and $SO(2)\times SU(2)\times SO(2)$ symmetry
is constructed in section \ref{SO2_SU2_SO2_Janus} while the
$N=(2,1)$ solution with $SO(2)\times SO(2)\times SO(2)$ symmetry
will be given in section \ref{SO2_3_Janus}. Conclusions and comments
on the results are given in section \ref{conclusions}. In the
appendix, we give a brief comments on the non-existence of
supersymmetric Janus solution with $SO(3)\times U(1)$ symmetry and
the possibility of other solutions within non-compact gauge groups.

\section{$N=3$ gauged supergravity}\label{N3theory}
Before giving the solutions, we review the $N=3$ gauged supergravity in four
dimensions and collect all relevant formulae which will be used in later sections. The reader is referred to
\cite{N3_Ferrara,N3_Ferrara2,Castellani_book} for the full construction. Apart from the mostly plus
metric signature $(-+++)$, all the notations are the same as in \cite{N3_Ferrara}.
\\
\indent In $N=3$ supersymmetry, the supergravity multiplet consists of the graviton $e^a_\mu$, three
gravitini $\psi_{\mu A}$, three vectors $A_{\mu A}$ and one spinor
field $\chi$. Indices $\mu,\nu,\ldots =0,\ldots,3$ and $a,b,\ldots=0,\ldots,3$ are respectively space-time and tangent space indices. The $SU(3)_R$ R-symmetry triplets are labeled by indices $A,B,\ldots=1,2,3$. Spinor indices will not be shown explicitly.
\\
\indent The only matter fields in $N=3$ supersymmetry are vector multiplets containing one vector field $A_\mu$, four spinors $(\lambda_{A},\lambda)$ which are a triplet and a singlet of $SU(3)_R$, and
three complex scalars $z_A$. Each vector multipet is labeled by indices $i,j,\ldots =1,\ldots, n$. Spinor fields are subject to the following chirality projections
\begin{eqnarray}
\psi_{\mu A}&=&\gamma_5\psi_{\mu A},\qquad \chi=\gamma_5\chi,\qquad
\lambda_A=\gamma_5\lambda_A,\qquad
\lambda=-\gamma_5\lambda,\nonumber \\
\psi_\mu^A&=&-\gamma_5\psi_\mu^A, \qquad
\lambda^A=-\gamma_5\lambda^A\, .
\end{eqnarray}
\indent The $N=3$ supergravity coupled to $n$ vector multiplets consists of $3n$ complex, $6n$ real, scalar fields $z_A^{\phantom{A}i}$ parametrized by the coset space
$SU(3,n)/SU(3)\times SU(n)\times U(1)$. The scalars can accordingly be
parametrized by the coset representative
$L(z)_\Lambda^{\phantom{\Lambda}\Sigma}$ which transforms under the
global $G=SU(3,n)$ and the local $H=SU(3)\times SU(n)\times U(1)$
symmetries by left and right multiplications, respectively.
\\
\indent Indices $\Lambda, \Sigma, \ldots=(A,i)=1,\ldots, n+3$ denote
fundamental representation of $SU(3,n)$. The coset representative
can also be decomposed into
$L_\Lambda^{\phantom{\Lambda}\Sigma}=(L_\Lambda^{\phantom{\Lambda}A},L_\Lambda^{\phantom{\Lambda}i})$.
The inverse of $L_\Lambda^{\phantom{\Lambda}\Sigma}$ is given in
term of $L_\Lambda^{\phantom{\Lambda}\Sigma}$ via the relation
\begin{equation}
(L^{-1})_\Lambda^{\phantom{\Lambda}\Sigma}=J_{\Lambda\Pi}J^{\Sigma\Delta}(L_\Delta^{\phantom{\Delta}\Pi})^*
\end{equation}
where $J_{\Lambda\Sigma}$ is an $SU(3,n)$ invariant tensor defined by
\begin{equation}
J_{\Lambda\Sigma}=J^{\Lambda\Sigma}=(\delta_{AB},-\delta_{ij}).
\end{equation}
\indent In the presence of $n$ vector multiplets, $(n+3)$-dimensional subgroups of $SO(3,n)\subset SU(3,n)$ can be gauged provided that its structure constants defined by the gauge algebra
\begin{equation}
\left[T_\Lambda,T_\Sigma\right]=f_{\Lambda\Sigma}^{\phantom{\Lambda\Sigma}\Gamma}T_\Gamma
\end{equation}
satisfy the consistency condition
\begin{equation}
f_{\Lambda\Sigma\Gamma}=f_{\Lambda\Sigma}^{\phantom{\Lambda\Sigma}\Gamma'}J_{\Gamma'\Gamma}
=f_{\left[\Lambda\Sigma\Gamma\right]}\, .
\end{equation}
A number of compact and non-compact gauge groups of this
``electric'' type have been studied in \cite{N3_4D_gauging}. In the
present work, we mainly focus on the case of $n=8$ vector multiplets
and compact $SO(3)\times SU(3)$ gauge group with the corresponding
structure constants given by
\begin{equation}
f_{\Lambda\Sigma}^{\phantom{\Lambda\Sigma}\Gamma}=(g_1\epsilon_{ABC},g_2
f_{i+3,j+3,k+3}),\qquad i,j=1,\ldots, 8\, .
\end{equation}
In the above equation, $f_{ijk}$ are the usual $SU(3)$ structure constants. In what follow, we are interested only in supersymmetric Janus solutions with only the metric and scalars non-vanishing. All the other fields will accordingly be omitted from the following discussion.
\\
\indent The bosonic Lagrangian of the $N=3$ gauged supergravity is given by
\begin{equation}
e^{-1}\mc{L}=\frac{1}{4}R-\frac{1}{2}P_\mu^{iA}P^\mu_{Ai}-V\, .
\end{equation}
The vielbein $P_i^{\phantom{i}A}$ of the $SU(3,n)/SU(3)\times
SU(n)\times U(1)$ coset are given by the $(A,i)$-components of the
Mourer-Cartan one-form
\begin{equation}
\Omega_{\Lambda}^{\phantom{\Lambda}\Pi}=(L^{-1})_\Lambda^{\phantom{\Lambda}\Sigma}dL_\Sigma^{\phantom{\Sigma}\Pi}
\end{equation}
with $\Omega_i^{\phantom{i}A}=(\Omega_A^{\phantom{A}i})^*$. The scalar
potential is given in terms of the ``boosted structure
constants''
\begin{equation}
C^\Lambda_{\phantom{\Lambda}\Pi\Gamma}=L_{\Lambda'}^{\phantom{\Lambda}\Lambda}
(L^{-1})_{\Pi}^{\phantom{\Lambda}\Pi'}(L^{-1})_{\Gamma}^{\phantom{\Lambda}\Gamma'}
f_{\Pi'\Gamma'}^{\phantom{\Pi'\Gamma'}\Lambda'}\qquad
\textrm{and}\qquad
C_\Lambda^{\phantom{\Lambda}\Pi\Gamma}=J_{\Lambda\Lambda'}J^{\Pi\Pi'}J^{\Gamma\Gamma'}
(C^{\Lambda'}_{\phantom{\Lambda}\Pi'\Gamma'})^*
\end{equation}
by the following relation
\begin{eqnarray}
V&=&-2S_{AC}S^{CM}+\frac{2}{3}\mc{U}_A\mc{U}^A+\frac{1}{6}\mc{N}_{iA}\mc{N}^{iA}
+\frac{1}{6}\mc{M}^{iB}_{\phantom{iB}A}\mc{M}_{iB}^{\phantom{iB}A}\nonumber \\
&=&\frac{1}{8}|C_{iA}^{\phantom{iA}B}|^2+\frac{1}{8}|C_i^{\phantom{A}PQ}|^2-\frac{1}{4}
\left(|C_A^{\phantom{A}PQ}|^2-|C_P|^2\right)
\end{eqnarray}
where $C_P=-C_{PM}^{\phantom{PM}M}$. All tensors appearing
in the above equations are defined by
\begin{eqnarray}
S_{AB}&=&\frac{1}{4}\left(\epsilon_{BPQ}C_A^{\phantom{A}PQ}+\epsilon_{ABC}C_M^{\phantom{M}MC}\right)\nonumber \\
&=&\frac{1}{8}\left(C_A^{\phantom{A}PQ}\epsilon_{BPQ}+C_B^{\phantom{A}PQ}\epsilon_{APQ}\right),\nonumber \\
\mc{U}^A&=&-\frac{1}{4}C_M^{\phantom{A}MA},\qquad \mc{N}_{iA}=-\frac{1}{2}\epsilon_{APQ}C_i^{\phantom{A}PQ},\nonumber \\
\mc{M}_{iA}^{\phantom{iA}B}&=&\frac{1}{2}(\delta_A^BC_{iM}^{\phantom{iM}M}-2C_{iA}^{\phantom{iA}B}).
\end{eqnarray}
\indent The supersymmetry transformations of fermionic fields are given by
\begin{eqnarray}
\delta \psi_{\mu A}&=&D_\mu \epsilon_A+S_{AB}\gamma_\mu \epsilon^B,\\
\delta \chi &=&\mc{U}^A\epsilon_A,\\
\delta \lambda_i&=&-P_{i\mu}^{\phantom{i}A}\gamma^\mu \epsilon_A+\mc{N}_{iA}\epsilon^A,\\
\delta
\lambda_{iA}&=&-P_{i\mu}^{\phantom{i}B}\gamma^\mu\epsilon_{ABC}\epsilon^C
+\mc{M}_{iA}^{\phantom{iA}B}\epsilon_B\, .
\end{eqnarray}
The covariant derivative on the supersymmetry parameter $\epsilon_A$
is defined by
\begin{equation}
D\epsilon_A=d\epsilon_A+\frac{1}{4}\omega^{ab}\gamma_{ab}\epsilon_A+Q_A^{\phantom{A}B}\epsilon_B+\frac{1}{2}nQ
\epsilon_A
\end{equation}
where the $SU(3)\times SU(8)\times U(1)$ composite connections $(Q_A^{\phantom{A}B},Q_i^{\phantom{i}j},Q)$ can be obtained from the $(A,B)$ and $(i,j)$ components of the Mourer-Cartan one-form via
\begin{equation}
\Omega_A^{\phantom{A}B}=Q_A^{\phantom{A}B}-n\delta^B_AQ,\qquad
\Omega_i^{\phantom{i}j}= Q_i^{\phantom{i}j}+3\delta^j_iQ\, .
\end{equation}
Note that $(Q_A^{\phantom{A}B},Q_i^{\phantom{i}j})$ satisfy $Q_A^{\phantom{A}A}=Q_i^{\phantom{i}i}=0$.
\\
\indent We can now construct the BPS equations for finding
supersymmetric Janus solutions. The metric ansatz takes the form of
$AdS_3$-sliced domain wall
\begin{equation}
ds^2=e^{2A(r)}\left(e^{\frac{2\xi}{\ell}}dx^2_{1,1}+d\xi^2\right)+dr^2\, .
\end{equation}
In the limit $\ell\rightarrow \infty$, this metric becomes a flat domain wall used in the study of holographic RG flows. The non-vanishing spin connections of the above metric can be computed to be
\begin{equation}
\omega^{\hat{\xi}}_{\phantom{\hat{\xi}}\hat{r}}=A'e^{\hat{\xi}},\qquad \omega^{\hat{\mu}}_{\phantom{\hat{\xi}}\hat{\xi}}=\frac{1}{\ell}e^{-A}e^{\hat{\mu}},\qquad
\omega^{\hat{\mu}}_{\phantom{\hat{\xi}}\hat{r}}=A'e^{\hat{\mu}}
\end{equation}
where $'$ denotes the $r$-derivative. From now on, indices $\mu,\nu$
will take values $0,1$, and hatted indices are the tangent space, or
flat, ones. In the above expressions, the vielbein components are
given by
\begin{equation}
e^{\hat{\mu}}=e^{A+\frac{\xi}{\ell}}dx^\mu,\qquad
e^{\hat{\xi}}=e^{A}d\xi,\qquad e^{\hat{r}}=dr\, .
\end{equation}
All of the scalar fields only depend on $r$. Therefore, only the $r$-component of $P^{iA}$ will be non-vanishing. The variations $\delta\lambda_i$ and $\delta\lambda_{iA}$ then require a $\gamma^r$ projection. Following \cite{warner_Janus}, this projection takes the form of
\begin{equation}
\gamma^{\hat{r}}\epsilon_A=e^{i\Lambda}\epsilon^A
\end{equation}
where $\Lambda$ is a real phase. Using the Majorana representation with all gamma matrices real and $\gamma_5=i\gamma_{\hat{0}}\gamma_{\hat{1}}\gamma_{\hat{\xi}}\gamma_{\hat{r}}$ purely imaginary, we have the relation $\epsilon^A=(\epsilon_A)^*$. This implies
\begin{equation}
\gamma^{\hat{r}}\epsilon^A=e^{-i\Lambda}\epsilon_A\, .
\end{equation}
\indent For the gravitino variations, we will denote the eigenvalues
of $S_{AB}$ matrix corresponding to the unbroken supersymmetry by
$-\frac{1}{2}\mc{W}$. $\mc{W}$ will play the role of the
``superpotential''. With this and the above spin connections, the
variation $\delta \psi_{A\hat{\mu}}=0$ gives
\begin{equation}
A'\gamma_{\hat{r}}\epsilon_A+\frac{1}{\ell}e^{-A}\gamma_{\hat{\xi}}\epsilon_{A}-\mc{W}\epsilon^A=0\, .\label{dPsi_mu_eq}
\end{equation}
As in \cite{warner_Janus}, taking the complex conjugate and
iterating the above equation lead to
\begin{equation}
A'^2=W^2-\frac{1}{\ell^2}e^{-2A}\label{dPsi_BPS_eq}
\end{equation}
where the ``real superpotential'' is defined by $W=|\mc{W}|$. We now
take the $\gamma_{\hat{\xi}}$ projection to be
\begin{equation}
\gamma_{\hat{\xi}}\epsilon_A=i\kappa e^{i\Lambda}\epsilon^A
\end{equation}
with $\kappa^2=1$.
\\
\indent The equation coming from $\delta \psi_{A\hat{\xi}}=0$ gives
\begin{equation}
e^{-A}\pd_\xi\epsilon_A+\frac{1}{2}A'\gamma_{\hat{\xi}\hat{r}}\epsilon_{A}
-\frac{1}{2}\mc{W}\gamma_{\hat{\xi}}\epsilon^A=0\, .
\end{equation}
Using equation \eqref{dPsi_mu_eq}, we find
\begin{equation}
\pd_\xi\epsilon_A=\frac{1}{2\ell}\epsilon_A
\end{equation}
which implies $\epsilon_A=e^{\frac{\xi}{2\ell}}\tilde{\epsilon}_A$
for $\xi$-independent $\tilde{\epsilon}_A$. Following
\cite{warner_Janus}, we will denote the Killing spinor by
\begin{equation}
\epsilon_A=e^{\frac{A}{2}+\frac{\xi}{2\ell}+i\frac{\Lambda}{2}}\varepsilon^{(0)}_A
\end{equation}
where the constant spinors $\varepsilon^{(0)}_{A}$ satisfy
\begin{equation}
\gamma_{\hat{r}}\varepsilon^{(0)}_{A}=\varepsilon^{(0)A}\qquad \textrm{and}\qquad \gamma_{\hat{\xi}}\varepsilon^{(0)}_{A}=i\kappa\varepsilon^{(0)A}\, .
\end{equation}
\indent In order to determine $e^{i\Lambda}$, we come back to equation \eqref{dPsi_mu_eq} and take the real and imaginary parts
\begin{eqnarray}
A'&=&\frac{1}{2}W(e^{i\omega-i\Lambda}+e^{-i\omega+i\Lambda}),\\
\frac{\kappa}{\ell}e^{-A}&=&\frac{i}{2}W(e^{i\omega-i\Lambda}-e^{-i\omega+i\Lambda})
\end{eqnarray}
where we have written $\mc{W}=We^{i\omega}$. There are two
possibilities namely real and complex $\mc{W}$. For real $\mc{W}$,
$\mc{W}=W$ or $\omega=0$, we find
\begin{equation}
e^{i\Lambda}=\frac{A'}{W}+\frac{i\kappa}{\ell}\frac{e^{-A}}{W}\, .\label{real_W_phase}
\end{equation}
For complex $\mc{W}$, we simply have
\begin{equation}
e^{i\Lambda}=\frac{\mc{W}}{A'+\frac{i\kappa}{\ell}e^{-A}}\, .\label{complex_W_phase}
\end{equation}
Both cases occur in the solutions considered in subsequent sections.

\section{Janus solution with $SO(2)\times SU(2)\times SO(2)$ symmetry}\label{SO2_SU2_SO2_Janus}
We begin with the solution with $SO(2)\times SU(2)\times SO(2)$
symmetry. There are two singlet scalars invariant under this
symmetry. They correspond to $SU(3,8)$ non-compact generators
\begin{equation}
\hat{Y}_1=e_{3,11}+e_{11,3} \qquad \textrm{and}\qquad
\hat{Y}_2=-ie_{3,11}+ie_{11,3}
\end{equation}
where we have used the matrices
$(e_{\Lambda\Sigma})_{\Gamma\Delta}=\delta_{\Lambda\Gamma}\delta_{\Sigma\Delta}$.
These are non-compact generators of $SU(1,1)\subset SU(3,8)$ with
the compact $U(1)$ subgroup generated by
\begin{equation}
J=2i(e_{33}-e_{11,11}).
\end{equation}
We can parametrize this $SU(1,1)/U(1)$ coset in the form of
\begin{equation}
L=e^{\varphi J}e^{\phi \hat{Y}_1}\, .
\end{equation}
The scalar potential can be computed to be
\begin{equation}
V=-\frac{1}{2}g_1^2[1+2\cosh(2\phi)]\label{potential}
\end{equation}
which admits only a critical point at $\phi=0$ as already studied in
\cite{N3_SU2_SU3}.
\\
\indent We now consider the BPS equations. With the above coset
representative and the $SO(3)\times SU(3)$ structure constants given
previously, we find the $S_{AB}$ matrix
\begin{equation}
S_{AB}=-\frac{1}{2}\textrm{diag}(\mc{W}_1,\mc{W}_1,\mc{W}_2)
\end{equation}
where
\begin{eqnarray}
\mc{W}_1&=&g_1\cosh(2\varphi)\cosh(\phi),\nonumber \\
\mc{W}_2&=&g_1e^{2i\varphi}\cosh\phi\, .
\end{eqnarray}
As pointed out in the RG flows studied of \cite{N3_SU2_SU3},
$\mc{W}_2$ gives rise to supersymmetric solutions. Using
$W=|\mc{W}_2|$, we can write the scalar potential as
\begin{equation}
V=-\frac{1}{8}\frac{1}{\sinh^2(2\phi)}\frac{\pd W}{\pd
\varphi}\frac{\pd W}{\pd \varphi}-\frac{1}{2}\frac{\pd W}{\pd
\phi}\frac{\pd W}{\pd \phi}-\frac{3}{2}W^2\, .
\end{equation}
\indent In order to solve all equations from the gravitini
variations, we set $\epsilon_{1,2}=0$, so the unbroken supersymmetry
is associated with $\epsilon_3$. The preserved supersymmetry for the
full solution will be $(1,0)$ or $(0,1)$ depending on the values of
$\kappa=1,-1$.
\\
\indent By computing $P^{iA}_r$ and using the projectors as
described in the previous section, we obtain two different complex
equations from $\delta\lambda_i=0$ and $\delta\lambda_{iA}=0$. It
turns out that the latter only involves $\epsilon_{1,2}$ and hence
identically vanishes. Solving for $\phi'$ and $\varphi'$, the BPS
equations from $\delta\lambda_i=0$, involving only $\epsilon_3$,
read
\begin{eqnarray}
\varphi'&=&\frac{\kappa}{4\ell}e^{-A}\textrm{sech}^2\phi,\label{Janus1_eq1}\\
\phi'&=&-\tanh\phi A'\, .\label{Janus1_eq2}
\end{eqnarray}
Together with the equation
\begin{equation}
A'^2-g_1^2\cosh^2\phi+\frac{e^{-2A}}{\ell^2}=0,\label{Janus1_eq3}
\end{equation}
we can solve for the supersymmetric solution. It can be verified
that these equations also solve the second-order field equations. In
the RG flow limit $\ell \rightarrow \infty$, we find
\begin{equation}
\varphi'=0,\qquad \phi'=\mp g_1\sinh\phi,\qquad A'=\pm g_1\cosh\phi
\end{equation}
which are the flow equations studied in \cite{N3_SU2_SU3}.
\\
\indent It is remarkable that equations \eqref{Janus1_eq1},
\eqref{Janus1_eq2} and \eqref{Janus1_eq3} turn out to be the same as
those considered in \cite{warner_Janus}. The solution can be
obtained similarly. By solving equation \eqref{Janus1_eq2}, we find
\begin{equation}
A=C_1-\ln\sinh\phi
\end{equation}
where $C_1$ is an integration constant.
\\
\indent Inserting the solution for $A$ into equation \eqref{Janus1_eq3}, we find an equation for $\phi$
\begin{equation}
\phi'^2=g_1^2\sinh^2\phi-\frac{e^{-2C_1}}{\ell^2}\frac{\sinh^4\phi}{\cosh^2\phi}\, .
\end{equation}
Follow \cite{warner_Janus}, we define a parameter
\begin{equation}
a=g_1\ell e^{C_1}\, .
\end{equation}
Accordingly, the solution for $\phi$ can be found to be
\begin{equation}
\sinh\phi=\zeta \frac{a}{\sqrt{1-a^2}}\frac{1}{\cosh[g_1(r-r_0)]}\label{Sol1_1}
\end{equation}
for $a<1$, and
\begin{equation}
\sinh\phi=\zeta \frac{a}{\sqrt{a^2-1}}\frac{1}{\sinh[g_1(r-r_0)]}\label{Sol1_2}
\end{equation}
for $a>1$. In these solutions, the parameter $\zeta=\pm 1$ can be
chosen to be $+1$ if we choose $\phi>0$. Furthermore, the
integration constant $r_0$ can be set to zero.
\\
\indent Using the solutions for $A$ and $\phi$ in equation \eqref{Janus1_eq1}, we obtain the solution for $\varphi$
\begin{eqnarray}
\tan (\varphi-\varphi_0)&=&-\kappa\zeta\sqrt{1-a^2}\sinh[g_1(r-r_0)],\qquad \textrm{for}\qquad a<1,\\
\tan (\varphi-\varphi_0)&=&-\kappa \zeta\sqrt{a^2-1}\cosh[g_1(r-r_0)],\qquad \textrm{for}\qquad a>1\, .
\end{eqnarray}
The metric warp factor can also be expressed as a function of $r$ as follow
\begin{eqnarray}
e^A&=&\zeta \frac{\sqrt{1-a^2}}{g_1\ell}\cosh[g_1(r-r_0)],\qquad \textrm{for}\qquad a<1,\\
e^A&=&\zeta \frac{\sqrt{a^2-1}}{g_1\ell}\sinh[g_1(r-r_0)],\qquad \textrm{for}\qquad a>1\, .
\end{eqnarray}
For further holographic study, it is useful to give an asymptotic
expansion of the solution near the $AdS_4$ critical point.
\\
\indent Similar to the discussion in \cite{warner_Janus}, the
solution with $a<1$ is smooth for $-\infty<r<\infty$ and approaches
the $SO(3)\times SU(3)$ $AdS_4$ critical point for $r\rightarrow \pm \infty$. Using the new radial
coordinate
\begin{equation}
r=\mp\frac{1}{g_1}\ln\left[\frac{\sqrt{1-a^2}}{2a}\rho\right],\label{rho1}
\end{equation}
we find that the limit $r\rightarrow \pm \infty$ corresponds to $\rho\rightarrow 0$. In this limit, the solution behaves as
\begin{eqnarray}
& &\phi\sim \rho+\frac{1}{12a^2}(a^2-3)\rho^3+\ldots,\nonumber \\
& &\varphi\sim \varphi_0\mp \kappa\frac{\pi}{2}\pm \kappa\frac{\rho}{a}\mp\kappa\frac{1+3a^2}{12a^3}\rho^3+\ldots,\nonumber \\
& & A\sim -\ln \rho+\ln\frac{a}{g_1\ell}+\frac{(1-a^2)}{4a^2}\rho^2+\ldots\label{Asymptotic1}
\end{eqnarray}
where we have set $\zeta=1$ and $r_0=0$. The leading terms simply give
\begin{equation}
A\sim \pm g_1r
\end{equation}
and
\begin{equation}
\phi\sim e^{\mp\frac{r}{L}},\qquad \varphi\sim e^{\mp\frac{r}{L}}
\end{equation}
where the $AdS_4$ radius is given by $L=\frac{1}{g_1}$. This indicates that $\phi$ and $\varphi$ are dual to relevant operators of dimensions $\Delta=1,2$ in the dual $N=3$ SCFT arising from the
$AdS_4\times N^{010}$ compactification.
\\
\indent As pointed out in \cite{Witten_operator_vev}, the values of
$\Delta=1,2$ can lead to two different quantizations.
Holographically, the two quantizations imply different
identifications of the operator deformations and vacuum expectation
values (vevs). In $N=8$ ABJM theory, it has been shown in
\cite{Warner_holo_super} that the correct holographic dictionary
requires the ``standard quantization'' for scalars and the
``alternative quantization'' for pseudoscalars. In the standard
quantization, non-normalizable modes are identified with
deformations while normalizable modes describe vevs. The
identification is reversed in the alternative quantization.
Consequently, the operators dual to scalars and pseudoscalars are
given respectively by bosonic and fermionic bilinears of dimensions
one and two. It would be interesting to determine whether there
exists such a unique dictionary in the case of $N=3$ SCFTs
considered here. This would make the holographic interpretation of
supergravity solutions more transparent.
\\
\indent For $a>1$, we still choose $\zeta=1$ and $r_0=0$ but define the new radial coordinate by
\begin{equation}
r=\mp \frac{1}{g_1}\ln\left[\frac{\sqrt{a^2-1}}{2a}\rho\right].\label{rho2}
\end{equation}
The behavior of $\phi$ and $\varphi$ near $r\rightarrow \pm \infty$
or $\rho\sim 0$ can be determined as in the previous case. The
result is the same as in \eqref{Asymptotic1}. Therefore, the
solution approaches the $AdS_4$ similar to the $a<1$ case.
\\
\indent However, the scalar $\phi$ and the metric function $A$
diverse at a finite value of $r=0$. There are two possibilities for
$r>0$ and $r<0$. For $r>0$, we choose $\zeta=1$, and for $r<0$ we
choose $\zeta=-1$ in order to make $e^A$ positive. We then find the
expansion near $|r|\approx 0$
\begin{eqnarray}
& &\phi \sim \mp\ln\left[\frac{\sqrt{a^2-1}g_1|r|}{2a}\right]+\ldots,\nonumber \\
& &\varphi\sim \mp\kappa\tan^{-1}\sqrt{a^2-1}+\ldots,\nonumber \\
& & e^A\sim \frac{\sqrt{a^2-1}}{\ell}|r|+\ldots\, .
\label{Asymptotic2}
\end{eqnarray}
It can be readily seen that $\phi$ and the metric become singular at
$r=0$.
\\
\indent From the scalar potential \eqref{potential}, we see that
$V(\phi\rightarrow\pm \infty)\rightarrow -\infty$. At least, by the
criterion of \cite{Gubser_singularity}, the singularity is
acceptable. It would be interesting to investigate this singularity
in eleven-dimensional context. The four-dimensional metric near this
singularity is given by
\begin{equation}
ds^2=\frac{a^2-1}{\ell^2}r^2ds^2(AdS_3)+dr^2\, .
\end{equation}
\indent The Janus solution for $a>1$ should accordingly correspond
to an interface between $N=3$ SCFT and a non-conformal field theory
or a Coulomb phase. As pointed out in \cite{ICFT_BCFT}, this
solution might also be useful in describing boundary conformal field
theories (BCFTs). In \cite{ICFT_BCFT}, it has been argued that the
strength of the deformation determines whether the solution
corresponds to an ICFT or a BCFT. In the present solution, the
deformation is determined by the parameter $a$. The fact that the
value of $a$ larger than a critical value gives the solution dual to
a BCFT is in agreement with the discussion in \cite{ICFT_BCFT}.

\section{Janus solution with $SO(2)\times SO(2)\times SO(2)$ symmetry}\label{SO2_3_Janus}
We now come to a more complicated solution with smaller residual
symmetry. There are four scalars invariant under $SO(2)\times
SO(2)\times SO(2)$ symmetry. They are given by the following
$SU(3,8)$ non-compact generators
\begin{eqnarray}
\tilde{Y}_1&=&e_{3,11}+e_{11,3},\qquad
\tilde{Y}_2=-ie_{3,11}+ie_{11,3},\nonumber \\
\tilde{Y}_3&=&e_{36}+e_{63},\qquad \tilde{Y}_4=-ie_{36}+ie_{63}\, .
\end{eqnarray}
These generators are also non-compact generators of $SU(2,1)\subset
SU(3,8)$. The corresponding scalars are then coordinates of a
submanifold $SU(2,1)/U(2)$. The resulting $SU(2,1)/SU(2)\times U(1)$
coset can be parametrized as in \cite{N3_SU2_SU3}
\begin{equation}
L=e^{\varphi_1 J_1}e^{\varphi_2 J_2}e^{\varphi_3 J_3}e^{\Phi
\tilde{Y}_1}
\end{equation}
where the generators $J_i$ form the $SU(2)$ subgroup. Explicitly,
they are given by
\begin{equation}
J_1=-ie_{66}+ie_{11,11},\qquad J_2=e_{6,11}-e_{11,6},\qquad
J_3=-ie_{6,11}-ie_{11,6}\, .
\end{equation}
The scalar potential takes a simple form
\begin{equation}
V=-\frac{1}{2}g_1^2[1+2\cosh(2\Phi)].
\end{equation}
\indent In this case, although the potential turns out to be the
same as in the previous case, the $S_{AB}$ matrix is given by
\begin{equation}
S_{AB}=-\frac{1}{2}g_1\cosh\Phi\delta_{AB}\, .
\end{equation}
We see that $S_{AB}$ has a three-fold degenerate real eigenvalue
giving rise to a real superpotential.
\\
\indent Under the $SO(2)_R\subset SO(3)_R$ identified with the first
$SO(2)$ in $SO(2)\times SO(2)\times SO(2)$, the supersymmetry
transformation parameters $\epsilon_A$ transform as
$\mathbf{2}+\mathbf{1}$. The singlet corresponds to $\epsilon_3$.
Similar to the previous cases, $\delta \lambda_i=0$ equations only
involve $\epsilon_3$ while $\delta\lambda_{iA}=0$ only have
non-vanishing components along $\epsilon_{1,2}$.
\\
\indent As pointed out in \cite{warner_Janus}, different
representations of $\epsilon_A$ under the residual symmetry can be
assigned different $e^{i\Lambda}$ phases. In the following, we will
choose the $\gamma_r$ projections to be
\begin{equation}
\gamma_{\hat{r}}\epsilon_{1,2}=e^{i\Lambda}\epsilon^{1,2}\qquad
\textrm{and}\qquad
\gamma_{\hat{r}}\epsilon_3=e^{-i\Lambda}\epsilon^3\, .
\end{equation}
For $\gamma_{\hat{\xi}}$ projection, we choose
\begin{equation}
\gamma_{\hat{\xi}}\epsilon_{1,2}=i\kappa e^{i\Lambda}\epsilon^{1,2}\qquad
\textrm{and}\qquad
\gamma_{\hat{\xi}}\epsilon_3=-i\kappa e^{-i\Lambda}\epsilon^3\, .
\end{equation}
This implies the opposite chirality of $\epsilon^{1,2}$ and
$\epsilon^3$ on the $(1+1)$-dimensional interface. With these
projectors and the expression for $e^{i\Lambda}$ given in
\eqref{real_W_phase}, $\delta \lambda_i=0$ and
$\delta\lambda_{iA}=0$ variations reduce to the same set of
equations
\begin{eqnarray}
4e^{2\Phi}\left[2g_1\sinh\Phi+ie^{i\Lambda}\sin(2\varphi_3)\sinh(2\Phi)\varphi_2'+2e^{i\Lambda}\varphi_3'\right]& &\nonumber \\
+2i(e^{4\Phi}-1)e^{i\Lambda}\cos(2\varphi_2)\cos(2\varphi_3)\varphi_1'&=&0,\\
8e^{2\Phi}\left[\sinh\Phi\left[g_1+ie^{i\Lambda}\cosh\Phi\sin(2\varphi_3)\varphi_2'\right]+e^{i\Lambda}\Phi'\right]& &\nonumber \\
+2i(e^{4\Phi}-1)e^{i\Lambda}\cos(2\varphi_2)\cos(2\varphi_3)\varphi_1'&=&0\, .
\end{eqnarray}
Note that choosing different projectors as shown above means that
equations from $\delta\psi_{A\mu}$ variations involving
$\epsilon_{1,2}$ and $\epsilon_3$ are complex conjugate of each
other. This is possible by the fact that $\mc{W}$ is real. Since in
this case $e^{i\Lambda}$ and $e^{-i\Lambda}$ differ effectively by a
sign change in $\kappa$ as can be seen from equation
\eqref{real_W_phase}, the solution to these equations then preserves
$N=(2,1)$ supersymmetry on the interface.
\\
\indent Solving all of these equations results in the following BPS equations
\begin{eqnarray}
\varphi_1'&=&\frac{2\kappa}{\ell}\frac{\cos\varphi_3\sec\varphi_2e^{-A}}{\cosh^2\Phi},\qquad
\varphi_2'=\frac{2\kappa}{\ell}\frac{\sin\varphi_3e^{-A}}{\cosh^2\Phi},\nonumber \\
\varphi_3'&=&-\frac{2\kappa}{\ell}\frac{\cos\varphi_3\tan\varphi_2e^{-A}}{\cosh^2\Phi},\qquad \Phi'=-A'\tanh\Phi,\nonumber\\
0&=&A'^2+\frac{e^{-2A}}{\ell^2}-g_1^2\cosh^2\Phi\, .
\end{eqnarray}
These equations can readily be verified to satisfy the corresponding field equations. It should also be noted that in the limit $\ell\rightarrow \infty$, we
recover the BPS equations for holographic RG flows studied in
\cite{N3_SU2_SU3}
\begin{equation}
\varphi_1'=\varphi_2'=\varphi_3'=0,\qquad \Phi'=\mp
g_1\sinh\Phi,\qquad A'=\pm g_1\cosh\Phi\, .
\end{equation}
From the above equations, $\Phi'$ and $A'$ equations form a close set
since they do not couple to all of the $\varphi_i$. These two
equations can be solved by the same solutions as in the previous
case. We will not give their explicit form here to avoid
repetitions.
\\
\indent With $\Phi$ and $A$ solutions as given in the previous
section, we can solve for $\varphi_i$, $i=1,2,3$, as follow
\begin{eqnarray}
\cos\varphi_3 &=&\zeta C_2\sec\varphi_2,\\
\tan(\varphi_1-\tilde{\varphi}_0)&=&\zeta\frac{2C_2\sin \varphi_2}{\sqrt{2-4C^2_2+2\cos(2\varphi_2)}},\\
\sinh[g_1(r-r_0)]&=&-\frac{\kappa}{\sqrt{1-a^2}}\tan\left[\frac{1}{2}\tan^{-1}
\frac{\tan(\varphi_1-\tilde{\varphi}_0)}{C_2}\right],\,\, a<1,\nonumber \\
\\
\cosh[g_1(r-r_0)]&=&-\frac{\kappa}{\sqrt{a^2-1}}\tan\left[\frac{1}{2}\tan^{-1}
\frac{\tan(\varphi_1-\tilde{\varphi}_0)}{C_2}\right],\,\, a>1.\nonumber \\
\end{eqnarray}
Note that for $\varphi_2=\varphi_3=0$, we recover the solutions of
previous section provided that the identifications $C_2=\zeta$ and
$\varphi_1=2\varphi$ are made. Therefore, turning on $\varphi_2$ and
$\varphi_3$ further breaks the $SO(2)\times SU(2)\times SO(2)$
symmetry to $SO(2)\times SO(2)\times SO(2)$, but enhanced
supersymmetry from $N=(1,0)$ to $N=(2,1)$.
\\
\indent We now briefly look at asymptotic behaviors of the solution.
In this case, the expansion is more complicated. Therefore, we will
only give the asymptotic behavior for $r\rightarrow \infty$. The
expansion for $r\rightarrow -\infty$ can be obtained similarly. By
defining the coordinate $\rho$ as in \eqref{rho1} and setting
$\zeta=1$ and $r_0=0$, we find
\begin{eqnarray}
\varphi_1\sim & &\tilde{\varphi}_0-\tan^{-1}\left[C_2\tan\frac{\kappa \pi}{12a^3}\right]+\frac{4\kappa C_2}
{a\left[1+C^2_2+(1-C^2_2)\cos\frac{\kappa\pi}{6a^3}\right]}\rho\nonumber \\
\qquad & & +\frac{8C_2(C_2^2-1)\sin\frac{\kappa\pi}{6a^3}}{a^2\left[1+C_2^2+(1-C_2^2)\cos\frac{\kappa\pi}{6a^3}\right]}\rho^2
+\ldots,\nonumber \\
\varphi_2\sim & &\alpha_0+\alpha_1\rho+\alpha_2\rho^2+\ldots,\nonumber \\
\varphi_3\sim & &\beta_0+\beta_1\rho+\beta_2\rho^2+\ldots
\end{eqnarray}
with the expansions for $\Phi$ and $A$ given by the $\phi$ and $A$ expansions in \eqref{Asymptotic1}. $\alpha_i$ and $\beta_i$ are constants depending on $C_2$ and $a$. Explicitly, $\alpha_i$ are given by
\begin{eqnarray}
\alpha_0&=&-\frac{\sqrt{1-C^2_2}}{12C_2^3}\left[5C_2^2-2+(2+7C_2^2)\cos\frac{\kappa\pi}{6a^3}\right]\sec^2\frac{\kappa\pi}{12a^3}\tan\frac{\kappa\pi}{12a^3} ,\nonumber \\
\alpha_1&=& \kappa\frac{\sqrt{1-C_2^2}}{2aC_2^3}\sec^4\frac{\kappa\pi}{12a^3}\left[C_2^2-2+(2+C_2^2)\cos\frac{\kappa\pi}{6a^3}\right],\nonumber \\
\alpha_2&=&\frac{\sqrt{1-C_2^2}}{a^2C_2^3}\left[6+C_2^2-(2+3C_2^2)\cos\frac{\kappa\pi}{6a^3}\right]\sec^4
\frac{\kappa\pi}{12a^3}\tan\frac{\kappa\pi}{12a^3}\, .
\end{eqnarray}
The explicit form of $\beta_i$ is much more complicated, so we
refrain from giving them here. It should be noted that the above
expansion reduces to the $SO(2)\times SU(2)\times SO(2)$ solution
for $C_2=\zeta=\pm 1$ with $\varphi_1\sim 2\varphi$ and
$\varphi_2\sim \varphi_3\sim 0$.
\\
\indent The solution for $a>1$ is singular as in the previous case.
The asymptotic expansion near $r\sim 0$ can be obtained analogously.
Near the singularity $r\sim 0$, the metric and $\Phi$ are singular
while all of the $\varphi_i$ remain finite. In this case, the
singularity also satisfies the criterion of
\cite{Gubser_singularity}. We then expect the solution to be dual to
a BCFT.

\section{Conclusions}\label{conclusions}
We have found supersymmetric Janus solutions within $N=3$ gauged
supergravity for $SO(3)\times SU(3)$ gauge group. The solutions have
$SO(2)\times SU(2)\times SO(2)$ and $SO(2)\times SO(2)\times SO(2)$
symmetries with unbroken $N=(1,0)$ and $(2,1)$ supersymmetry on the
$(1+1)$-dimensional interface. The solutions provide a holographic
dual of conformal interfaces in the $N=3$ Chern-Simons-Matter theory
with $SU(3)$ flavor symmetry in three dimensions and might be
interesting in applications to condensed matter physics systems
along the line of \cite{N3_and_QHE}. Similar to the maximal $N=8$
theory, Janus solutions require non-vanishing pseudoscalars as
opposed to the RG flow solutions. In the case of $SO(3)\times U(1)$
symmetry in which the BPS equations require constant pseudoscalars,
there does not exist supersymmetric Janus solutions. This will be
shown in the appendix.
\\
\indent It would be very interesting to identifies precisely the
interface SCFTs dual to the gravity solutions given here. Since the
solutions given here are all analytic, they could be useful in a
holographic study of the correlation functions in the dual $N=3$
dCFT by the method introduced in \cite{Skenderis_correlation}. It
would be interesting to further study the solutions that become
singular in the IR, at a finite value of the $AdS_4$ radial
coordinate and give a precise interpretation in the dual field
theory. According to \cite{ICFT_BCFT}, these solutions should be
interpreted as gravity dual of boundary conformal field theories
(BCFTs). The analytic solutions should also be of particular
interest in this context as well as in computations of the
entanglement entropy.
\\
\indent Another interesting direction is obviously to look for
possible uplift of these solutions to eleven dimensions and identify
the corresponding M-brane configurations. The full embedding of
$N=3$ gauged supergravity considered in this paper to eleven
dimensions is currently not known. This is due to the lack of a
complete reduction ansatz on $N^{010}$ although an embedding keeping
only $SU(3)$ singlets has been constructed in
\cite{N010_truncation_Cassani}. Since the solutions found in the
present paper involve scalars that transform non-trivially under
$SU(3)$, a more general truncation is needed.
\\
\indent On the other hand, the full embedding might be possibly
first in $N=4$ gauged supergravity. As has been pointed out in some
previous works, see for example
\cite{Shadow_N3_multiplet,N010_truncation_Cassani,Jerome_IIB_to_N4_5D},
the $N=3$ $AdS_4$ vacuum can be realized as a supersymmetry breaking
$AdS_4$ vacuum of $N=4$ gauged supergravity. It would be desirable
to explicitly construct this truncation and study the
eleven-dimensional uplift of the Janus solutions found here and the
holographic RG flows in \cite{N3_SU2_SU3} similar to the solutions
of the maximal gauged supergravity recently studied in
\cite{Warner_N8_uplift}.
\vspace{0.5cm}\\
{\large{\textbf{Acknowledgement}}} \\
This work is supported by Chulalongkorn University through
Ratchadapisek Sompoch Endowment Fund under grant GF-58-08-23-01 (Sci-Super II). The author would like to thank Henning Samtleben, Nicolay Bobev and Davide Cassani for useful correspondences.

\appendix
\section{Supersymmetric solutions with $SO(3)\times U(1)$ symmetry}
In this appendix, we consider supersymmetric solutions with
$SO(3)\times U(1)$ symmetry and look for possible Janus solutions.
It turns out that, in this case, there is no supersymmetric Janus
solution. We now present some details of the analysis.
\\
\indent It has been found in \cite{N3_SU2_SU3} that apart from the
$SO(3)\times SU(3)$ symmetric $AdS_4$ critical point, there is
another $N=3$ critical point with $SO(3)\times U(1)$ symmetry. The
unbroken $SO(3)$ is a diagonal subgroup of $SO(3)\times SO(3)$ with
the second factor being $SO(3)\sim SU(2)\subset SU(2)\times
U(1)\subset SU(3)$. The $U(1)$ is an explicit $U(1)$ factor in
$SU(2)\times U(1)$. The uplift of this critical point to eleven
dimensions is presently unknown.
\\
\indent Under $SO(3)_{\textrm{diag}}\times U(1)$, there are two singlet scalars corresponding to the non-compact generators
\begin{eqnarray}
\hat{Y}_1&=&e_{14}+e_{41}+e_{25}+e_{52}+e_{36}+e_{63},\nonumber \\
\hat{Y}_2&=&-ie_{14}+ie_{41}-ie_{25}+ie_{52}-ie_{36}+ie_{63}\, .
\end{eqnarray}
The coset representative can be parametrized by
\begin{equation}
L=e^{\varphi J}e^{\Phi \hat{Y}_1}e^{-\varphi J}\, .
\end{equation}
where
\begin{equation}
J=\textrm{diag}(2i\delta^{AB},-2i\delta^{i+3,j+3},0,0,0,0,0),\qquad
i,j=1,2,3\, .
\end{equation}
The $J$ generator, corresponding to $U(1)\sim SO(2)$, together with $\hat{Y}_1$ and $\hat{Y}_{2}$ form a non-compact group $SU(1,1)\subset SU(3,8)$. The $SO(3)\times U(1)$ critical point is given by
\begin{equation}
\varphi=0,\qquad \Phi=\frac{1}{2}\ln
\left[\frac{g_2-g_1}{g_2+g_1}\right],\qquad
V_0=-\frac{3g_1^2g_2^2}{2(g_2^2-g_1^2)}\, .
\end{equation}
\indent We now consider the BPS equations for supersymmetric solutions. The scalar matrix $S_{AB}$ takes the form of
\begin{equation}
S_{AB}=-\frac{1}{2}\mc{W}\delta_{AB}
\end{equation}
where
\begin{eqnarray}
\mc{W}&=&\frac{1}{8}e^{-3\Phi}\left[\left[(1+e^{2\Phi})^3g_1+(e^{2\Phi}-1)^3g_2\right]\cos(2\varphi)
\right.\nonumber
\\
&
&\left.-i\left[(1+e^{2\Phi})^3g_1-(e^{2\Phi}-1)^3g_2\right]\sin(2\varphi)\right].
\end{eqnarray}
The variations $\delta\lambda_i$ and $\delta\lambda_{iA}$ give the
following equations
\begin{equation}
\Phi'-\frac{1}{3}e^{-i\Lambda}\frac{\pd \mc{W}}{\pd \Phi}\pm
ie^{-2\Phi}(e^{4\Phi}-1)\varphi'=0
\end{equation}
which implies $\varphi'=0$ or $\varphi=\varphi_0$. We will set the
constant $\varphi_0=0$ in order to recover the two $AdS_4$ critical
points at $r\rightarrow \pm \infty$. Using the expression for
$e^{i\Lambda}$ in term of $\mc{W}$, equation
\eqref{complex_W_phase}, we find that the above equation requires
\begin{equation}
\Phi=0\qquad \textrm{or}\qquad \Phi=\frac{1}{2}\ln \left[\frac{g_2-g_1}{g_2+g_1}\right].
\end{equation}
This means the scalars are fixed at the critical points.
\\
\indent The metric function $A(r)$, for $\Phi=0$, can be determined
from the gravitini variations
\begin{equation}
 A'^2+\frac{e^{-A}}{\ell^2}-g_1^2=0
\end{equation}
whose solutions are
\begin{eqnarray}
A&=&\ln\left[\frac{e^{-g_1r}(e^{g_1r}+g_1^2\ell^2)^2}{4g_1^4\ell^4}\right]\\
\textrm{or}\qquad A&=&\ln\left[\frac{e^{-g_1r}(e^{g_1r}g_1^2\ell^2+1)^2}{4g_1^4\ell^4}\right].
\end{eqnarray}
\indent For $\Phi=\frac{1}{2}\ln\left[\frac{g_2-g_1}{g_2+g_1}\right]$, we find
\begin{equation}
e^{A}\ell^2(g_2^2-g_1^2)A'^2=g_1^2(1+e^Ag_2^2\ell^2)-g_2^2\, .
\end{equation}
With a suitable integration constant, the solutions can be written as
\begin{eqnarray}
e^A&=&\frac{e^{-\frac{g_1g_2}{\sqrt{g_2^2-g_1^2}}r}}{16g_1^2g_2^2\ell^4}\left[e^{\frac{g_1g_2}{\sqrt{g_2^2-g_1^2}}r}
+4\ell^2(g_2^2-g_1^2)\right]^2\\
\textrm{or}\qquad
e^A&=&\frac{e^{-\frac{g_1g_2}{\sqrt{g_2^2-g_1^2}}r}}{16g_1^2g_2^2\ell^4}\left[4\ell^2(g_2^2-g_1^2)e^{\frac{g_1g_2}{\sqrt{g_2^2-g_1^2}}r}
+1\right]^2.
\end{eqnarray}
\indent These solutions are nothing but $AdS_4$ backgrounds in the
$AdS_3$-sliced parametrization. Therefore, there are no
supersymmetric Janus solutions connecting the $SO(3)\times U(1)$
critical point identified in \cite{N3_SU2_SU3}.
\\
\indent In addition, we have also looked for Janus solutions in the
case of non-compact gauge groups. It has been pointed out in
\cite{N3_4D_gauging} that among the ``electric'' gaugings, $SO(3,1)$
and $SL(3,\mathbb{R})$ gauge groups admit supersymmetric $AdS_4$
vacua. In the case of $SO(3,1)$, there is no supersymmetric Janus
solution with $SO(3)$ symmetry. For $SL(3,\mathbb{R})$ gauge group,
there are no scalars which are singlets under $SO(3)\subset
SL(3,\mathbb{R})$. Another possibility would be to consider
solutions with $SO(2)\subset SO(3)$ symmetry.
\\
\indent In both gauge groups, there are six $SO(2)$ singlets
parametrized by the coset manifold $SU(1,1)/U(1)\times
SU(1,1)/U(1)\times SU(1,1)/U(1)$ which is a submanifold of
$SU(3,3)/SU(3)\times U(3)$ and $SU(3,5)/SU(3)\times U(5)$,
respectively. The analysis turns out to be highly complicated. In
addition, there does not seem to be any simple truncation to fewer
scalars that can give rise to supersymmetric Janus solutions. It
would be useful to carry out the full analysis and definitely
determine whether $N=3$ gauged supergravity with $SO(3,1)$ and
$SL(3,\mathbb{R})$ gauge groups admits supersymmetric Janus
solutions.


\end{document}